\documentclass{PoS}

\title{Mass-radius constraints for compact stars and a critical endpoint}

\ShortTitle{M-R constraints for a CEP}

\author{\speaker{David Blaschke}\\
        Institute for Theoretical Physics,
University of Wroc{\l}aw,
50-204 Wroc{\l}aw, Poland\\
	Fakult\"at f\"ur Physik,
	Universit\"at Bielefeld,
	33501 Bielefeld, Germany
\\
Bogoliubov  Laboratory of Theoretical Physics,
JINR Dubna, 141980 Dubna, Russia
\\
        E-mail: \email{blaschke@ift.uni.wroc.pl}}

\author{David E. Alvarez-Castillo\\
        Bogoliubov  Laboratory of Theoretical Physics,
JINR Dubna, 141980 Dubna, Russia\\
Instituto de F\'{\i}sica,
Universidad Aut\'onoma de San Luis Potos\'{\i},
S.L.P. 78290, M\'exico
\\
        E-mail: \email{alvarez@theor.jinr.ru}
}

\author{Sanjin Beni\'c\\
        Physics Department, Faculty of Science, University of Zagreb,
Zagreb 10000, Croatia\\
        E-mail: \email{sanjinb@phy.hr}
}

\abstract{   
We present two types of models for hybrid compact stars composed of a quark 
core and a hadronic mantle with an abrupt first order phase transition at the 
interface which are in accordance with the latest astrophysical measurements 
of two 2 M$_\odot$ pulsars. 
While the first is a schematic one, the second one is based on a QCD motivated 
nonlocal PNJL model with density-dependent vector coupling strength.
Both models support the possibility of so called \textit{twin} compact stars 
which have the same mass but different radius and internal structure
at high mass ($\sim$ 2 M$_\odot$), provided they exhibit a large 
jump $\Delta \epsilon$ in the energy density of the first order phase 
transition fulfilling $\Delta \epsilon/\epsilon_{\rm crit} > 0.6$. 
We conclude that the measurement of high-mass twin stars would support the 
existence of a first order phase transition in symmetric matter at zero 
temperature entailing the existence of a critical end point in the QCD phase 
diagram.
}

\FullConference{8th International Workshop on Critical Point and Onset of 
Deconfinement,\\
		March 11 to 15, 2013\\
		Napa, California, USA}

\begin{document}

\section{Introduction}

The quest for the location of the critical endpoint (CEP) in the QCD 
phase diagram (PD) in heavy-ion collisions (HIC) is ongoing and has not yet 
yielded definite results. 
Suspicions are raised that there might be none(!) since that transition which
is seen to be crossover at vanishing chemical potential $\mu=0$ on the 
temperature axis of the PD (lattice QCD results of the Wuppertal-Budapest
collaboration \cite{Borsanyi:2010cj} and the HotQCD Collaboration 
\cite{Bazavov:2011nk} may remain of this type also at $\mu \neq 0$.
Such a behaviour has been obtained in local Polyakov-Nambu-Jona-Lasinio (PNJL)
models with a vector meson coupling \cite{Redlich:2006rf,Sasaki:2006ws}.
In particular, when this coupling is fixed so that the slope of the 
pseudocritical temperature obtained in lattice QCD simulations 
\cite{Kaczmarek:2011zz} at small $\mu$ is reproduced on the meanfield level of 
description \cite{Bratovic:2012qs}.
In nonlocal generalizations of the PNJL model the situation is not so clear;
it depends on the details of the mdel setup 
\cite{Contrera:2012wj,Hell:2012da}.  
However, as understanding confinement within these models is not possible or 
questionable and also the backreaction of hadronic fluctuations on the phase 
structure is not accomplished, no firm satements about the location and even 
the existence of the CEP in the PD can be made at present.

This situation is rather unsatisfactory on the background of 
large scale experimental programmes to experimentally identify signatures of 
the CEP and identify its location which are operating at RHIC Brookhaven (STAR
BES programme) and CERN-SPS (NA61) and are in preparation 
(NICA at JINR Dubna and CBM at FAIR Darmstadt).

Therefore, at this workshop an appeal was issued to the community of compact 
star astrophysicists to come up with suggestions how to prove the existence of 
a first order QCD phase transition in compact star interiors.
Would it be possible to gain evidence for a strong first order 
deconfinement transition in compact stars this would prove the existence 
of at least one CEP in the QCD phase diagram since in the high temperature 
region explored by lattice QCD simulations  \cite{Kaczmarek:2011zz}
the transition is crossover.

One confirmative answer to the quest for a first-order phase transition in 
compact stars will be discussed here.
If nature is so kind let cold, high-density compact star matter undergo a 
strong first-order phase transition which fulfills certain conditions to
be detailed here, we show that the resulting mass-radius diagram for compact 
stars will provide besides an almost vertical hadronic star branch also a 
disconnected, hybrid star branch. If the onset of quark deconfinement, marking 
the end of this hadronic branch will occur at the presently known upper limit
of compact star masses, around $2~M_\odot$, then the ``third family'' of 
compact stars, the quark core hybrid stars, will occur also in this mass range
provided the quark matter EoS is stiff enough to carry the hadronic mantle.

What would it require to prove that this situation occurs in nature? 
As we know already two of these high-mass compact stars:
PSR J1614-2230 with 1.97 $\pm $ 0.04 M$_{\odot}$ \cite{Demorest:2010bx}
and PSR J0348+0432 with 2.01 $\pm $ 0.04 M$_{\odot}$ \cite{Antoniadis:2013pzd},
the measurement of their radii to an accuracy of about 5\% 
(i.e., $\delta~R \sim 600$m) bears the potential proof.
If their radii would differ significantly, e.g., by 2~km, whie their masses 
are about the same, they would present an example of the high-mass twin stars,
the existence of which would prove the presence of a strong first order 
phase transition in old compact stars and thus the existence of a CEP in the 
QCD phase diagram.   

\section{Massive hybrid stars \& twins}

In the following we want to demonstrate on the examples of two classes of 
hybrid EoS models under which conditions the interesting phenomenon of twins
at high compact star masses of $\sim 2~M_\odot$ may be obtained. 
The first one is a phenomenological ansatz  Zdunik, Haensel 
\cite{Zdunik:2012dj}, Alford, Han and Prakash \cite{Alford:2013aca}
which we call the ZHAHP scheme. 
The second one is based on a QCD motivated, microscopic EoS obtained within
a nonlocal Polyakov-NJL model, see \cite{Blaschke:2013rma} and references
therein. 

\subsection{The ZHAHP scheme}
A first order phase transition in neutron star matter can take place just as 
in symmetric matter where it is searched for in heavy ion collisions. 
Adopting the setting of the ZHAHP scheme, we construct 
hybrid stars with a hybrid EoS composed of a given hadronic EoS, here 
DD2~\cite{Typel:1999yq}, and a quark matter EoS parametrized by its 
squared speed of sound $c_{\rm QM}^2$ which pretty well describes 
\cite{Zdunik:2012dj} results of a color superconducting NJL 
model~\cite{Lastowiecki:2011hh}
\begin{equation}
\label{eos}
P(\varepsilon)= 
P_{\rm DD2}(\varepsilon)\Theta(\varepsilon_{\rm crit}-\varepsilon)
+c_{\rm QM}^2~\varepsilon~\Theta(\varepsilon-\varepsilon_{\rm crit}
-\Delta\varepsilon)~.
\end{equation}
The critical energy density $\varepsilon_{\rm crit}$ and the discontinuity 
$\Delta\varepsilon$ complete this three-parameter EoS model which is capable 
of describing compact star sequences with a third family of stars 
in the mass-radius diagram. 
For early works on the disconnected, third branch of stable compact stars and 
the related mass-twin phenomenon, see Refs. 
\cite{Gerlach:1968zz,Kampfer:1981yr,Schertler:2000xq,Glendenning:2000gh}.
Searching for sequences with twins obeying the $2~M_\odot$ mass constraint 
\cite{Demorest:2010bx,Antoniadis:2013pzd} we obtain a
quasi-horizontal hybrid star branch disconnected by an unstable branch from 
the almost vertical hadron star branch, as a consequence of a strong phase 
transition. 

\begin{figure}[!ht]
\includegraphics[width=0.5\textwidth,angle=-90]{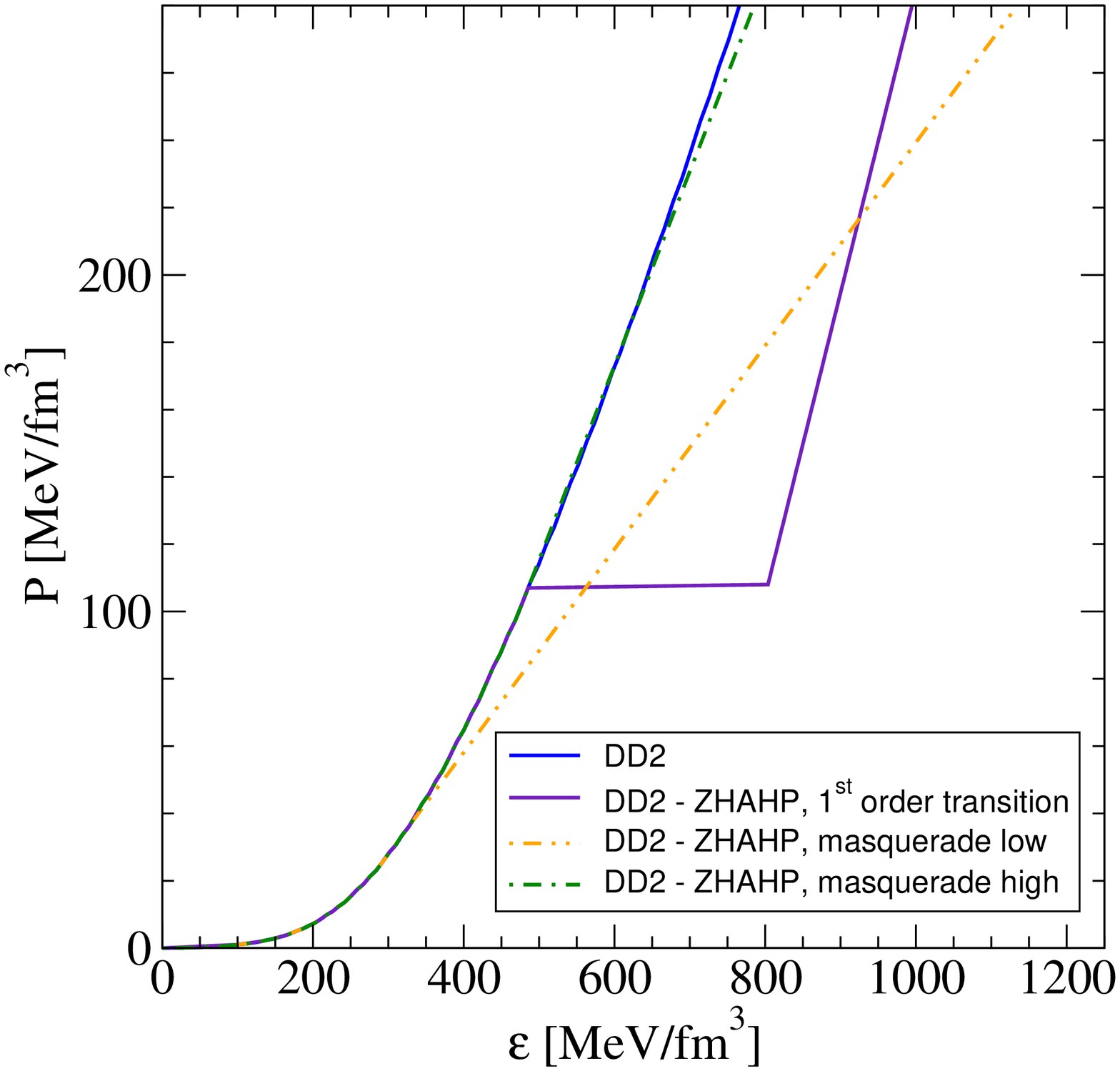}
\hspace{-2cm}
\includegraphics[width=0.5\textwidth,angle=-90]{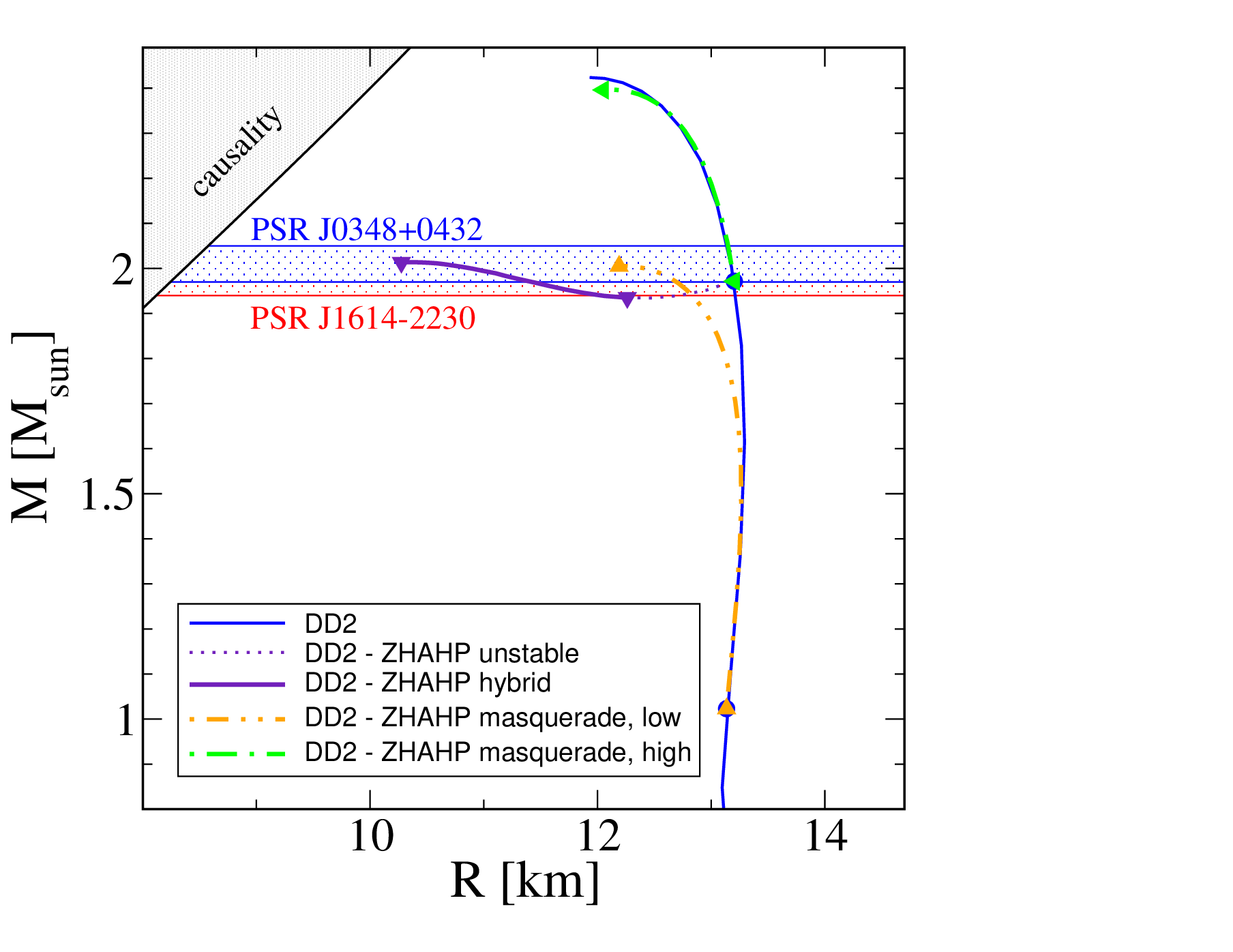}
\caption{\label{EoS-MR1}
Left panel: Hadronic EoS (DD2, blue solid line) vs. hybrid one in the ZHAHP 
scheme of Eq.~(\protect\ref{eos}) with a strong first order transition 
(indigo solid line) and with a crossover behaviour ($\Delta\varepsilon=0$)
at a low critical density ($\varepsilon_{\rm crit}=487$ MeV,
orange dash-double dotted line) and at a high one 
($\varepsilon_{\rm crit}=290$ MeV, green dash-dotted line).
Right panel:
Mass-radius diagram for hybrid star EoS in the ZHAHP scheme 
(\protect\ref{eos}). 
The almost vertical branch of pure hadronic stars (blue solid line) 
is separated from the almost horizontal hybrid star branch (indigo solid line) 
by an unstable branch (indigo dashed line) for a strong first order transition 
to stiff high-density matter. 
For comparison, we show two crossover situations, at low density 
(orange dash-double dotted line) and at high density (green dash-dotted line). 
The triangles mark the limits of the stable hybrid star branches.
For parameters, see text.
Recently measured masses for two 2 M$_\odot$ pulsars shown by red and blue 
hatched regions.}
\end{figure}

Fig.~\ref{EoS-MR1} (left) shows the corresponding EoS (\ref{eos}) for the 
parameters: 
$c_{\rm QM}^{2}= 0.94$, $\Delta\varepsilon = 0.67~\varepsilon_{\rm crit}$ and 
$\varepsilon_{\rm crit}=485$ MeV/fm$^3$. 
The latter corresponds to $P(\varepsilon_{\rm crit})=100$ MeV/fm$^3$ and a 
baryon density at the quark matter onset of 
$n_{\rm crit}= 2.9~n_0$ with $n_0=0.16$ fm$^{-3}$. 
Fig.~\ref{EoS-MR1} (right) shows the mass-radius relation for this hybrid EoS. 

The ZHAHP scheme allows also to study the dependence of the $M-R$ relation on 
the strength of the transition, i.e. on the parameter $\Delta\varepsilon$ 
which can be also set to zero, thus mimicking the situation of a crossover 
transition. When hadronic and high-density phases behave rather similar, this 
case is also called ``masquerade'' \cite{Alford:2004pf}. 
In Fig.~1, we show the EoS (left) and the $M-R$ relation (right) in this
situation for the two cases: a transition at a low density 
$n_{\rm crit}=1.8~n_0$ corresponding to $\varepsilon_{\rm crit}=290$ MeV, 
$c_s=0.55$; and at a high density $n_{\rm crit}=2.9~n_0$ corresponding to 
$\varepsilon_{\rm crit}=487$ MeV, $c_s=0.76$. 

\subsection{Nonlocal PNJL model}

\begin{figure}[!ht]
\includegraphics[width=0.5\textwidth,angle=-90]{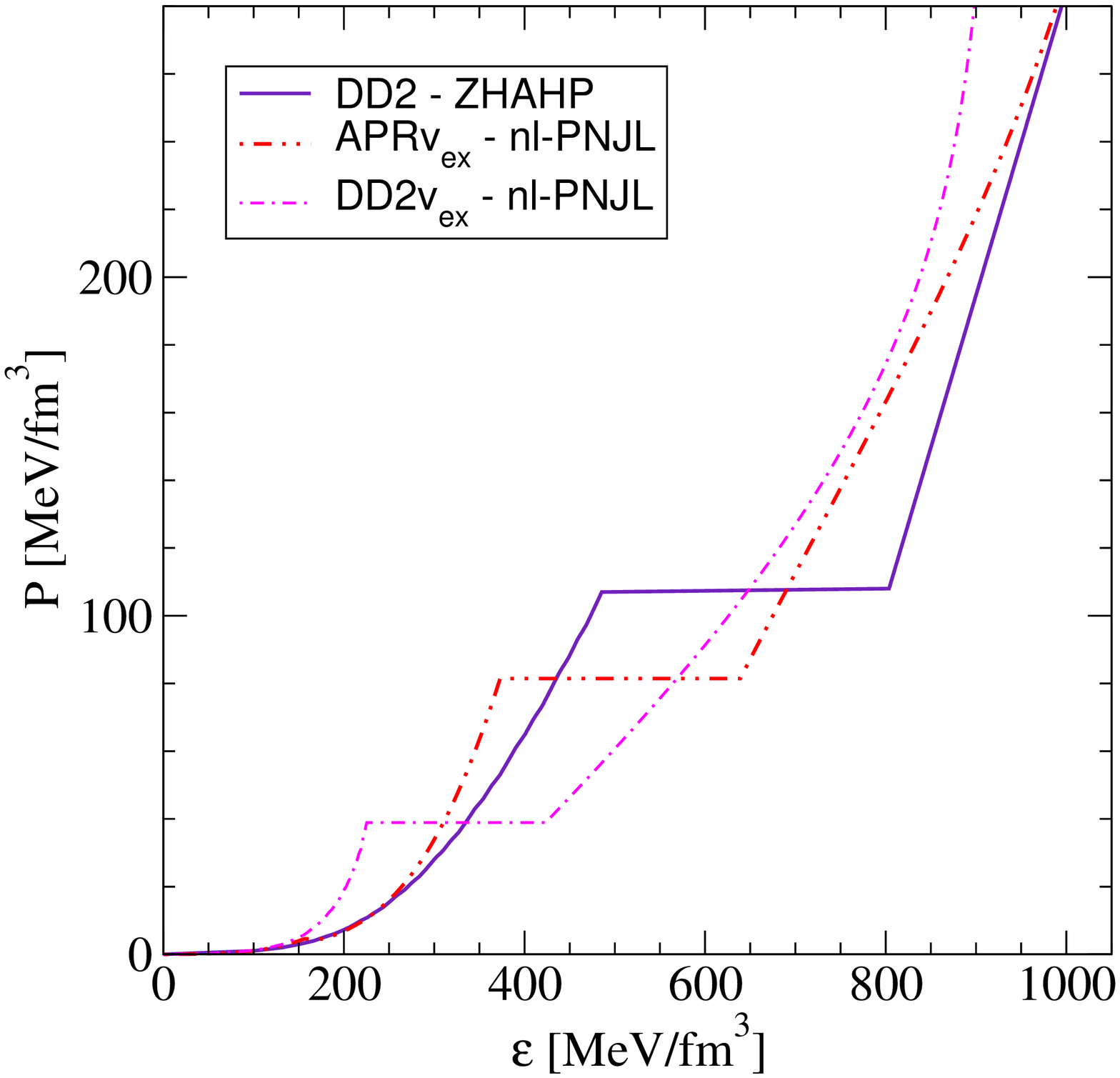}
\hspace{-2cm}
\includegraphics[width=0.5\textwidth,angle=-90]{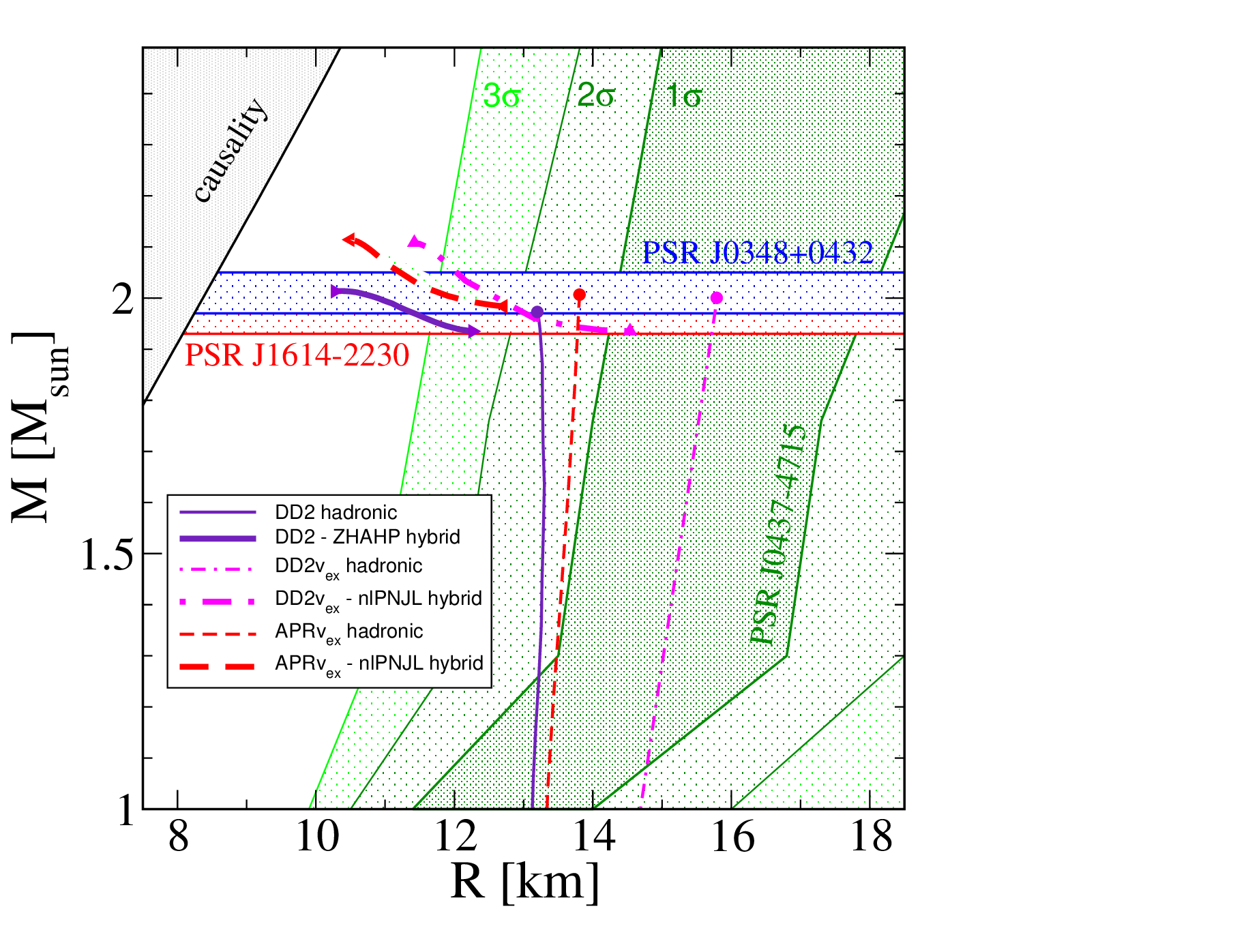}
\caption{\label{EoS-MR2}
Left panel: Hadronic EoS (DD2, green dashed line) vs. hybrid one 
(Eq.~\protect\ref{eos}, indigo solid line).
Right panel:
Mass-radius diagram for hybrid star EoS (\protect\ref{eos}). 
The almost vertical branch of pure hadronic stars (green solid line) 
is separated from the almost horizontal hybrid star branch (indigo solid line) 
by an unstable branch (black dashed line).
Recently measured masses for two 2 M$_\odot$ pulsars shown by red and blue 
hatched regions.}
\end{figure}

In this microscopic scheme for obtaining high-mass twins we use the standard
two-phase construction of the deconfinement phase transition, adopting separate
EoS models for the $\beta-$ equilibrated $T=0$ compact star matter:
$P_H(\mu)$ for the hadronic phase and $P_Q(\mu;\eta_v)$ for the quark phase. 
For the former we employ standard nuclear matter EoS like APR 
\cite{Akmal:1998cf} and DD2 \cite{Typel:1999yq} 
(see also \cite{Typel:2005ba,Klahn:2006ir}) 
which we modify at high densities by adopting an excluded volume with a 
nonlinear dependence on the chemical potential  \cite{Sasaki:2013mha}
\begin{equation}
v_{\rm ex}(\mu) = (4\pi/3) r^3(\mu)~~,~~r^3(\mu)=r_0 + r_1 \mu + r_2 \mu^2~.
\end{equation}
For the quark phase, we use the nonlocal PNJL model EoS of 
Refs.~\cite{Contrera:2012wj,Benic:2013eqa} where the 4-momentum dependence of 
the formfactors is adjusted to describe the dynamical mass function and wave 
function renormalization of the $T=0$ quark propagator from lattice QCD 
simulations \cite{Parappilly:2005ei,Kamleh:2007ud}
and the vector meson coupling $\eta_v$ is adjusted to obtain the
slope of the chemical potential dependence of the pseudocritical temperature
$T_c(\mu)$ in accordance with lattice QCD \cite{Kaczmarek:2011zz}.
In addition, we follow the procedure suggested in  \cite{Blaschke:2013rma}
to implement a $\mu$ dependence of $\eta_v$ by interpolating between zero
temperature pressures $P_< =P_Q(\mu;\eta_<)$ and $P_> =P_Q(\mu;\eta_>)$ 
in $\beta-$ equilibrium which are calculated at different, but fixed values 
$\eta_< = \eta_v(\mu \le \mu_c)$ and $\eta_> = \eta_v(\mu > \mu_c)$.
The critical chemical potential $\mu_c$ is found from the Gibbs condition for
the phase equilibrium $P_H(\mu_c)=P_<(\mu_c)$, where $P_H(\mu)$ stands for
the pressure of one of the hadronic EoS models in $\beta-$equilibrium.
Different from  \cite{Blaschke:2013rma} we use here a Gaussian function for 
this interpolation and obtain for the resulting hybrid star matter EoS
\begin{eqnarray}
\label{interpol}
P(\mu)&=&P_H(\mu)\Theta(\mu_c-\mu)+\left[\left(P_< - P_>\right)
\exp\left[-(\mu-\mu_c)^2/\Gamma^2\right]+P_>\right]\Theta(\mu-\mu_c)~.
\end{eqnarray}
The density jump at the first order phase transition is given by the change 
in the slope of the pressure at the critical chemical potential
$\Delta n = \partial P_</\partial \mu |_{\mu=\mu_c}
- \partial P_H/\partial \mu |_{\mu=\mu_c}$.
Note that with the choice $\eta_< < \eta_>$ one achieves a strong first order
phase transition with large $\Delta n$ on the one hand and stable hybrid star
configurations even at high central energy densities on the other. 
In particular, one can realize in this way microscopically motivated hybrid
star EoS which also exhibit the feature of mass twin star configurations at
high masses $M\sim 2 M_\odot$. 
Two particular examples are shown in Fig.~\ref{EoS-MR2} compared to the 
ZHAHP scheme parametrization of Ref.~\cite{Alvarez-Castillo:2013cxa}, given 
also in Fig.~\ref{EoS-MR1}. 
These examples show that it is possible to prove the presence of a strong 
first order phase transition in compact star matter provided nature would be 
so kind to allow the mass twin phenomenon to occur. 

The two parameter sets are given in Table~\ref{tab:par} 
\begin{table}
\begin{tabular}{l|c|c|c|c|c|c|c|c}
\hline
&$r_0$&$r_1$&$r_2$&$\varepsilon_{\rm crit}$&$\mu_c$&$\Gamma$&$\eta_<$&$\eta_>$
\\
model&[fm]&[fm/GeV]&[fm/GeV$^2$]&[MeV/fm$^3$]&[MeV]&[MeV]& &\\
\hline
DD2v$_{\rm ex}$ - nlPNJL&0.3&0.35&0.05&225.3&1420&360&0.10&0.20\\
APRv$_{\rm ex}$ - nlPNJL&0.65&0.01&0.0&373.0&1400&400&0.05&0.25\\
\hline
\end{tabular}
\caption{Parameters of the hybrid EoS model. For details, see text.}
\label{tab:par}
\end{table}
One may ask for the motivation of the increasing stiffness of quark matter
with chemical potential. This is a generally open question and should be
answered within a fully nonperturbative QCD approach to dense matter in the
vicinity of the deconfinement transition, which is not available yet.
So we may speculate that at asymptotically high chemical potentials the 
ratio of vector to scalar coupling can be given by the value from Fierz 
transformation of one-gluon exchange, $\eta_v=0.5$. 
Approaching the regio of the phase transition from above, the hadronic 
correlations might lead to a lowering of this value, eventually to as low 
values as we use here, $\eta_v(\mu_c)=0.05 ... 0.10$.
%Note also that in Dyson-Schwinger equation approaches at finite chemical 
%potentials, there is a tendency for the chiral condensate to grow with $\mu$.

\section{Conclusions}

In this work we have demonstrated for a simple hybrid star EoS model in the 
ZHAHP scheme as well as for more elaborated, microscopically motivated models,
that a strong first order phase transition in cold nuclear matter under 
neutron star constraints reveals itself by the mass twin phenomenon in the 
mass-radius diagram for compact star sequences.
Particularly interesting for verification by observations is the case of 
high-mass twins at $M\sim 2 M_\odot$. 
As our examples show, those twins have typically a difference in their radii of
about $2$ km. 
A further precondition for the existence of high-mass twins is a large radius
for massive stars on the hadronic branch, as indicated in some recent analyses 
\cite{Trumper:2011,Bogdanov:2012md}.
Therefore, if radius measurements of compact stars in that mass range could be 
performed to an accuracy of less than 1 km, this would allow in principle to 
detect such high-mass compact star twins, if they exist.
Their detection in turn would yield important impact to studies of the QCD 
phase diagram since the proof of a strong first order phase transition
at $T=0$ would imply the existence of a line of critical endpoints at finite 
temperatures.
Then, there can be hope to find signatures of a strong first-order phase 
transition also in heavy-ion collision experiments with cool, strongly 
compressed baryon matter. 
A very concrete suggestion derived from this study would be an observational 
campaign to measure the radii of the known $2~M_\odot$ pulsars, 
PSR J1614-2230 and PSR J0348+0432, for instance with the planned missions 
LOFT \cite{Mignani:2012vc} and/or NICER.

\section{Acknowledgements}
We gratefully acknowledge numerous discussions and collaboration work on the
topic addressed in this contribution with our colleagues, in particular with 
G. Contrera, H. Grigorian, O. Kaczmarek, T. Kl\"ahn, E. Laermann, 
R. {\L}astowiecki, M. C. Miller, G. Poghosyan, S. B. Popov, J. Tr\"umper, 
D. N. Voskresensky and F. Weber.
This research has been supported by Narodowe Centrum Nauki within the 
``Maestro'' programme under contract number DEC-2011/02/A/ST2/00306.
D.B. acknowledges support by the Russian Fund for Basic Research under grant 
number 11-02-01538-a. D.A-C. is grateful for support by the Bogoliubov-Infeld 
programme.

\end{document}